\documentclass[10pt,twocolumn,twoside]{article}
\usepackage{hyperref}
\usepackage{authblk}
\usepackage[switch]{lineno}
\usepackage{abstract}
\usepackage{epstopdf}
\usepackage{graphicx}
\usepackage{comment}
\usepackage[normalem]{ulem}
\usepackage{xcolor}
\usepackage[round,authoryear]{natbib}
\usepackage{amsmath,amssymb}
\usepackage{enumitem}
\usepackage{setspace}

\title{Evolution as fitness landscape navigation: concepts, measures, and emerging questions}

\author[1]{Malvika Srivastava}
\author[2,3]{Claudia Bank}
\author[4]{Joachim Krug}
\author[2,3]{Suman G. Das\thanks{Corresponding author. Email: \texttt{sgdas.work@gmail.com}}}

\affil[1]{Department of Ecology and Evolution, University of Lausanne, Lausanne, Switzerland}
\affil[2]{Institute of Ecology and Evolution, University of Bern, Bern, Switzerland}
\affil[3]{Swiss Institute of Bioinformatics, Lausanne, Switzerland}
\affil[4]{Institute for Biological Physics, University of Cologne, Cologne, Germany}

\begin{document}
\nolinenumbers

\twocolumn[
\maketitle
\begin{onecolabstract}
Fitness landscapes are mappings between genotypes, phenotypes, and fitness that shape evolution. In recent years, empirical work and theoretical models have greatly advanced our understanding of how populations navigate rugged fitness landscapes. Here, we provide a timely review of the theoretical aspects of this field. Its rapidly growing literature employs a wide range of terms, which are sometimes used ambiguously or inconsistently. We therefore begin by defining the major concepts and the field's vocabulary, highlighting our own terminology choices wherever needed. We then review key results on the relationships between epistasis, ruggedness, accessibility, and navigability for genotype-fitness maps, highlighting several complex and sometimes counterintuitive connections that have emerged. Further, we review how the conserved structural properties of the underlying genotype-phenotype map, which can lead to the formation of large connected neutral networks of genotypes, influence dynamics on fitness landscapes. We then compare the two levels to study landscape navigation: the level of genotype-phenotype maps and the level of genotype-fitness maps. Our review leads us to propose a new measure of navigability, based on evolutionary outcomes, that is broadly applicable and overcomes limitations of existing measures. Finally, we highlight examples from the smaller body of work that relaxes the common assumption of fitness-monotonic paths on static landscapes, and discuss how this can fundamentally change the nature of fitness landscape navigation. Throughout the review, we identify directions for future work to fill existing gaps and to synthesize the disparate strands of research within the field.
\end{onecolabstract}
]





\section{Introduction}

\begin{figure}[htbp]
\centering
\includegraphics[width=235pt]{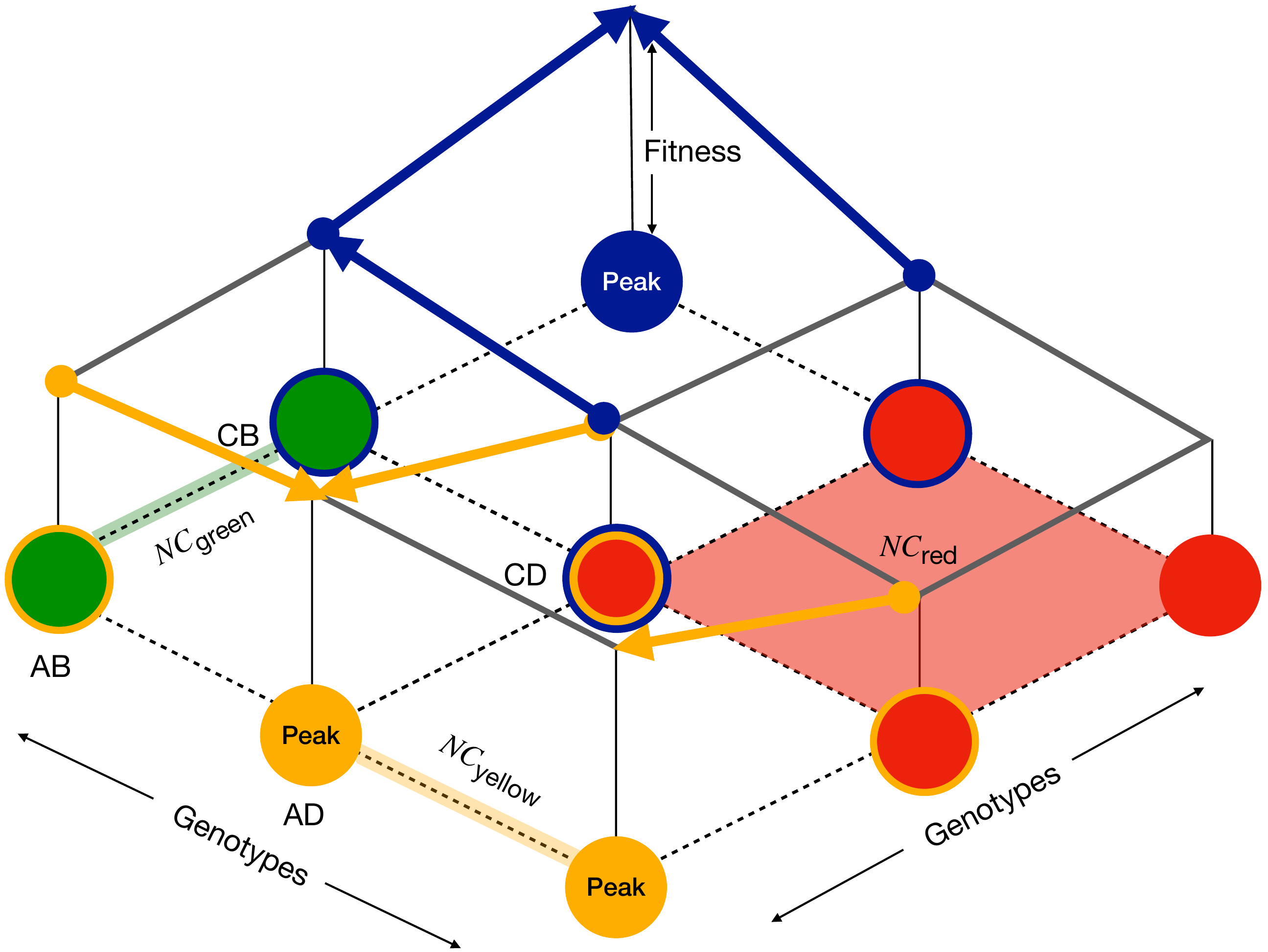}
\caption{\textbf{Schematic diagram of a genotype-phenotype-fitness landscape.} Circles represent genotypes, and dashed edges connect genotypes to their \textit{genotypic neighbors}. The interior colour of each genotype represents its phenotype; together, the circles and their colours define a \textit{genotype--phenotype (GP) map}. The vertical elevation of each genotype corresponds to its fitness, and these elevations collectively define the genotype--fitness map, or the \textit{fitness landscape}. The genotypes $AB$, $AD$, $CD$, and $CB$ exhibit \textit{sign epistasis}, since the mutation from $B \to D$ is beneficial on the $A$ background but deleterious on the $C$ background. Coloured arrows indicate \textit{accessible paths}: they point in the direction of increasing fitness, and their colour denotes the peak phenotype at which the walk terminates. The \textit{adaptive basin} of each peak (here blue and yellow) is highlighted by the border 
colour of the genotype circles belonging to that basin. Neutral mutations are shown as dark grey undirected edges. The red genotype $CD$ has a \textit{genotypic robustness} of $\frac{1}{2}$. In a greedy walk, a population at this genotype proceeds to the yellow peak genotype $AD$. However, in a random \textit{adaptive walk}, the population may instead move to the green genotype $CB$ and ultimately reach the blue peak. The genotypes at the vertices of the red square form the largest \textit{neutral component (NC)}, here of size 4. The \textit{phenotypic evolvability} of the NC corresponding to the red phenotype, $\epsilon_{\text{red}}$, is 3, since three distinct phenotypes (blue, green, and yellow) can be reached by single mutational steps from the red NC. Similarly, $\epsilon_{\text{green}} = 3$, $\epsilon_{\text{yellow}} = 2$, and $\epsilon_{\text{blue}} = 2$. For a less condensed illustration, separate schematics of the genotype--phenotype map and the fitness landscape are provided in Supplementary Figure~\ref{sfig:1}.
}\label{fig:1}
\end{figure}

The concept of fitness landscapes as maps from genotypes to fitness was introduced by \citet{wright1932roles}, accompanied by a classic metaphorical image that depicted a rugged terrain with multiple peaks and valleys. For a long time, the fitness landscape served primarily as a metaphor for understanding evolution as a selection-driven process of ``climbing'' fitness peaks \citep{svensson2012adaptive}. Recent decades, however, have experienced a renewed interest in fitness landscapes driven by empirical results emphasizing the ubiquity of \textit{epistasis} (reviewed in \citealp{phillips2008epistasis,de2014empirical,domingo2019causes}), and by the technological possibility to measure empirical fitness landscapes (reviewed in \citealp{de2014empirical},\citealp{bank2022epistasis} and \citealp{Ogbunugafor2026}).  In addition, an influx of ideas from other disciplines, such as statistical physics \citep{stein1992spin, stadler2002fitness}, computer science \citep{richter2014recent}, and complex systems \citep{kauffman1995home, stadler2007towards}, has introduced new theoretical models and methods for quantifying and predicting epistatic effects.

Recent years have seen an especially steep increase in studies on landscape navigation, which aim to understand how a population subject to the forces of evolution finds its way to high-fitness regions of complex fitness landscapes. A key term in this context is the ``navigability" of a fitness landscape.
To our knowledge, this term -- now common in the literature -- was introduced into modern fitness landscape theory less than a decade ago by \citet{aguilar2017thousand}. The authors defined landscape navigability as ``the ability to access a global peak via an evolutionary exploration involving random mutation
and natural selection". This work was soon followed by a study of genotype-phenotype (GP) maps \citep{manrubia2017distribution} where navigability was defined as ``the ease with which alternative phenotypes are reached". While the term is relatively new, the topic itself is much older, and some of the key questions go back all the way to \citet{wright1932roles}. What has changed rapidly in recent years is the availability of increasingly large fitness landscape datasets and high computational power, facilitating empirically-informed answers to questions that were once posed in the abstract. Thus, the navigation of fitness landscapes is no longer just a metaphor for evolution. 

Here, we take the opportunity to review this growing field of theory, which complements the steadily growing body of empirical fitness landscape studies \citep[reviewed in][]{de2014empirical, fragata2019evolution, bank2022epistasis, ascensao2026experimental}. We restrict ourselves to theory that conceptualizes fitness landscapes as mappings from a discrete genotype space (i.e., a hypercube or \textit{Hamming space}) to phenotypes or fitness, and that aims to quantify the navigation process across the whole landscape in the long term. This explicitly excludes theoretical approaches from quantitative genetics \citep{walsh2018evolution} and adaptive dynamics \citep{Avila2023}, which   have been used to address related questions. As we highlight throughout the paper, the ``shape" of such mappings from genotype to phenotype or fitness, which ultimately determines the ruggedness and navigability of the fitness landscape, is crucially determined both by the topology of the genotype space, i.e., the connectedness and size of the 
Hamming space, and the topography of the mapping, i.e. the values to which each genotype is mapped. 

Also within our restricted selection of literature and due to the interdisciplinarity of this research area, we perceive a large variation in approaches and viewpoints to the study of fitness landscapes, some of which we would like to highlight here. Studies of probabilistic fitness landscapes (section 3A, B) often focus on local or global statistics of the topography of the fitness landscape and how these may yield analytical predictions regarding navigation of a given landscape. This focus could stem from the field's vicinity to statistical physics, which
has lead to a long-standing exchange of methods and concepts 
\citep{kauffman1987towards,stein1992spin,hwang2018universality,das2022driven}. The study of the speed of adaptation on probabilistic fitness landscapes (section 3A) has been taken up by computer scientists interested in search problems on complex graphs \citep{kaznatcheev2019computational}.
In comparison, studies of genotype-phenotype maps (section 3C, D) are often directly motivated by complex empirical data or numerical models of RNA folding, resulting in a focus on simulations and on studying specific trajectories towards a ``target phenotype". Here, optimization is invoked not necessarily with regard to fitness but with regard to keeping options open while remaining ``stable" \citep{Manrubia_2021,Greenbury2022NavigableGP}. Further, the possibility for neutral plateaus and genotypic redundancy generates fitness landscapes with varying topologies.
Finally, the original questions posed by \citet{wright1932roles} came from evolutionary biology, a field in which the relative importance of epistatic and beneficial selection has been continually debated. Notably, Wright argued in his seminal paper that {\em because} of his proposed ruggedness of the fitness landscape, other evolutionary forces such as migration or genetic drift are crucial to carry populations across fitness peaks. Therefore, we discuss in section 4 how shifting the focus away from (strong) selection and weak mutation as the main evolutionary processes during landscape navigation may take the field forward.   

Given its interdisciplinarity and rapid growth, terms related to the study of fitness landscapes are sometimes used in an inconsistent or ambiguous manner. As noted before, the term ``navigability'' itself was introduced differently by two different studies. To address this challenge, we begin our review with a glossary of terms.
Importantly, throughout the paper, we distinguish between the terms ``navigability'' and ``navigation''. We use the latter term to refer to the process of moving across a landscape under the influence of evolutionary forces. 
The term ``navigability'' -- similar to the term ``ruggedness'' -- does not have a consensus quantitative definition, but  refers to some feature of the fitness landscape that describes the ``ease'' of evolutionary movement across it. Properties of landscape topography such as ruggedness -- quantified, for example, by the number of fitness peaks -- have been used as indicators of navigability. However, the question of biological interest is the reachability of high-fitness peaks in a landscape, and based on our review below, we argue that the topographic measures used in the literature are not, in general, adequate indicators of this. Ultimately, we propose a new measure called ``navigational ruggedness'' which is based on evolutionary outcomes and which, we argue, serves as a generally informative measure of navigability. 

\section{Glossary}
This section contains an alphabetical glossary of terms defining fundamental concepts and measures used in the field and featured in this review. Alternative or uncommon usages of terms in the literature are noted where this helps avoid confusion. For ease of reference, terms that appear in the glossary are italicized in the text. 

\begin{description}[itemindent=0pt,leftmargin=0pt]

\item[Accessible path] A {\em path} is a succession of single-mutation steps that connect two genotypes. A (selectively) {\em accessible path} is a path for which every step is fitness-increasing \citep{weinreich2005perspective,franke2011evolutionary}. A path is called {\em direct} if its length (number of steps) is equal to the Hamming distance between the pair of genotypes and {\em indirect} otherwise. A genotype $g^\prime$ is \textit{accessible} from genotype $g$ when there exists an accessible path from $g$ to $g^\prime$. Accessible paths are indicated by thick blue and yellow arrows in Figure \ref{fig:1}. In the context of navigation of GP maps, the definition includes paths with neutral steps (shown with grey lines).

\item[Adaptive basin] The adaptive basin of a peak is the set of all genotypes from which accessible paths to this peak exist. Adaptive basins of the peaks are highlighted with thick coloured borders in Figure \ref{fig:1}. An adaptive basin is sometimes called the {\em basin of attraction}. We do not use this term to avoid confusion with gradient basins.

\item[Adaptive walk]  An adaptive walk provides a description of the evolutionary process in the limit of strong selection and weak mutation (SSWM regime, see 
\citealp{gillespie1984molecular}, \citealp{orr2002population} and \citealp{mccandlish2014modeling}). Formally, it is a Markov chain that traces an accessible path. In other words, at every step of the adaptive walk, the genotype is updated to a fitter neighboring genotype. In a random adaptive walk, the new genotype is chosen uniformly at random from among all fitter genotypic neighbors \citep{kauffman1987towards}.  A greedy or gradient walk is a deterministic limiting case in which the new genotype is always the fittest neighboring genotype \citep{orr2003minimum}. 

\item[Allele graph] A graph showing all possible allelic states at each
locus in a genotype with edges connecting mutational neighbors
\citep{SchmiegeltKrug2023}.

\item[Alphabet cardinality] The number of possible alleles at each locus
in a genotype; e.g., it is 4 (nucleotides) for DNA sequences and
20 (amino acids) for protein sequences. 

\item[Epistasis] The dependence of the fitness effect of a mutation on the genetic background in which it occurs \citep{phillips2008epistasis,poelwijk2016context,domingo2019causes,johnson2023epistasis}.

\item[Evolvability of a genotype-phenotype (GP) map] In the GP map literature, evolvability, as a measure of adaptive potential, has been defined at two different levels. At the genotypic level, (genotypic) evolvability is the number of different phenotypes that can be reached by mutational changes in a given genotype, whereas at the phenotypic level, (phenotypic) evolvability is the number of different phenotypes that can be reached by mutational changes in the genotypes that encode the given phenotype (see the caption of Figure \ref{fig:1} for the phenotypic evolvability of the different phenotypes) \citep{wagner2008robustness}. 

\item[Fitness landscape] In this review, a fitness landscape is a map from a discrete genotype space to fitness (or to a phenotype used as a fitness proxy) (see Figure \ref{fig:1}).

\item[Genotype-phenotype (GP) map] A map that specifies how genetic variation produces phenotypic variation (see Figure \ref{fig:1}). A GP map can be defined at any biological scale, for instance, the mapping between RNA sequences and their secondary structure or the mapping between a set of genes and the body-plan morphology they produce.

\item[Genotypic neighbor] Sometimes simply called a `neighbor', a genotypic neighbor is a genotype at Hamming distance $d=1$ from the reference genotype, that is, it differs from the reference genotype at a single locus (see Figure \ref{fig:1}).

\item[Global epistasis] A simplified form of epistasis where the dependence on genetic background reduces to a dependence on background fitness or some other low-dimensional phenotype \citep{diaz2023global}. In the context of protein evolution
this is also referred to as non-specific epistasis \citep{domingo2019causes}.

\item[Gradient basin] The gradient basin (or greedy basin) of a peak is the set of genotypes that can reach the peak through a greedy adaptive walk \citep{stadler2010combinatorial}. Note that gradient basins of two peaks cannot overlap, whereas \textit{adaptive basins} can. 

\item[Hamming space] A graph with nodes representing genotype sequences of length $L$ with symbols drawn from an alphabet of $a$ letters, and links
representing point mutations \citep{stadler2002fitness}. The Hamming space of binary sequences ($a=2$) is the $L$-dimensional hypercube.

\item[Navigability] The navigability of a fitness landscape is the ease with which an evolving population can move across it. It does not have a consensus quantitative definition but is estimated in the literature in various ways (see subsequent sections). In the specific context of the GP map literature, navigability of a genotype-phenotype-fitness map is defined as the average probability of existence of an accessible path between pairs of phenotypes, and \textit{evolutionary navigability} is defined as the average probability that an adaptive walk reaches a target phenotype from a source phenotype via an accessible path\citep{Greenbury2022NavigableGP}.

\item[Neutral Component (NC)] A maximal set of mutationally connected genotypes mapping to the same phenotype \citep{Greenbury_2016}, see Figure \ref{fig:1} for example. 

\item[Random field models of fitness landscapes] Probabilistic models that assign random variables to genotypes \citep{stadler1999random}. Random field models have become a standard tool for benchmarking empirical data and exploring generic properties of fitness landscapes.  Examples include the following:
\begin{itemize}
\item[(i)] The House-of-Cards (HoC) or mutational landscape model \citep{gillespie1984molecular,kauffman1987towards}, where fitness is assigned randomly to each genotype.
\item[(ii)] $NK$ models \citep{kauffman1989nk,hwang2018universality}, where fitness is the sum of contributions of $L$ loci, and the contribution of a locus depends both on its state and the state of $K$ other loci with which it interacts. The  value of the fitness contribution is chosen randomly. The $NK$ model reduces to the HoC model for $K=L-1$.    
\item[(iii)] Rough Mount Fuji (RMF) model
\citep{aita2000analysis,szendro2013quantitative,Neidhart_2014}, where the fitness of a genotype is the sum of additive contributions from each locus and a random term specific to the genotype. The strength of the random term can be tuned to modify the degree of epistasis in the landscape. 
\item[(iv)] Fisher's geometric model \citep{blanquart2014properties, hwang2017genotypic}, where fitness is a single-peaked function {defined on a continuous Euclidean} phenotype space, and mutational effects on phenotypes are additive. 
\end{itemize}

\item[Reachability] In this paper, we have defined the reachability of any genotype as the probability of reaching that genotype through an adaptive walk starting from specified initial conditions \citep{das2020predictable}. Unless otherwise mentioned, the initial genotype is chosen randomly. 

\item[Robustness] Genotypic robustness is the fraction of genotypic neighbors of a given genotype that have the same phenotype (see Figure \ref{fig:1} for example). Phenotypic robustness is the average 
genotypic robustness of all genotypes that encode the given phenotype \citep{wagner2008robustness}.

\item[Ruggedness] A fitness landscape is called rugged 
when it deviates from additivity of fitness effects, particularly through sign epistasis or 
if it has multiple local fitness peaks. 

\item[Sign epistasis] occurs when the sign of a mutation (beneficial or deleterious) depends on the genetic background \citep{weinreich2005perspective}. For example, in Figure \ref{fig:1} the mutation from $B \to D$ at the second locus is beneficial when the first locus carries an allele $A$, but is deleterious when the first locus carries an allele $C$. Reciprocal sign epistasis (RSE) occurs when mutations at two loci exhibit sign epistasis relative to each other \citep{poelwijk2007empirical}. 

\end{description}

\section{Navigation of fitness landscapes through accessible paths}

\subsection{Epistasis and shape of fitness landscapes: overview}
Given a genotype space, the topography of a \textit{fitness landscape} determines the background on which evolutionary processes unfold and can thus inform the ease and predictability of evolution. The simplest topography is that of an additive landscape, where the effect of multiple mutations is equal to the sum of their individual effects. Such a landscape is single-peaked, and all steps toward the peak are fitness-increasing, making the peak easy to reach. Thus, in additive landscapes, the evolutionary endpoint is predictable, although the specific mutational path leading to it is not. \textit{Epistasis} refers to interactions between mutational effects, and landscapes with such interactions can be 
{\textit{rugged}}, potentially reducing both the ease and predictability of evolution. Therefore, to understand the ease and predictability of adaptation at different biological scales, it is important to ascertain whether empirical fitness landscapes at different scales are smooth or rugged. 

Several measures of ruggedness have been proposed and studied \citep{carneiro2010adaptive,szendro2013quantitative,de2014empirical,ferretti2016measuring,aguilar2017thousand}. Early comparative studies of relatively small empirical landscapes found that different measures of ruggedness were generally consistent \citep{szendro2013quantitative,de2014empirical}. One of the most commonly used measures of ruggedness is the number of peaks in a landscape. It is naturally relevant in the study of fitness landscape navigation since questions often focus on the likelihood of reaching the global peak (or a comparatively high peak in the landscape). A key concern regarding the navigation process, already formulated by \cite{wright1932roles}, is that a population may become trapped on local, low-fitness peaks when a landscape is highly rugged. 

The relationship between pairwise (i.e., local) epistasis and the number of peaks is complex. Classic work by \cite{poelwijk2011reciprocal} proved that the existence of a reciprocal \textit{sign epistasis} (RSE) motif is necessary for the existence of multiple peaks. The result was generalized by \cite{saona2022relation} using discrete Morse theory to show that $n$ peaks require the existence of at least $n-1$ instances of RSE. The same result, but specifically for di-allelic loci, was obtained by a different method in \cite{riehl2022occurrences}. Although RSE is necessary for the existence of multiple peaks, it is not sufficient. Strikingly, the results in \cite{riehl2022occurrences} imply that in the limit of a large number of loci, it is possible to have a single-peaked landscape where almost any randomly picked pair of loci in any genetic background exhibits RSE.

Results on the sufficient conditions for the existence of multiple peaks remain limited. \cite{crona2013peaks} showed that a system with instances of RSE but no other sign epistasis must have at least two peaks, but no sufficient condition for the existence of general $n$ peaks is known. It has been argued that information on `local' properties based on lower-order interactions, such as the existence of a few RSE motifs in the landscape, is not likely to provide much information on a global property such as the number of peaks. Motivating arguments for this view, based on Morse theory, were provided in \cite{poelwijk2011reciprocal}, but a systematic theoretical development of these ideas has been lacking.

The results above involving RSE have been based mostly on landscapes without neutral mutations. Recent work \citep{ivankov2026sign} has shown that when neutral mutations are allowed, pairwise sign epistasis can be absent from multi-peaked landscapes, but a modified RSE based on composite mutations necessarily exists. 

Fitness peaks are interesting as end-points of \textit{adaptive walks}, whereas the layouts of \textit{accessible paths} show how epistasis can guide the evolution of populations toward fitness peaks. \cite{weinreich2005perspective} brought into focus the importance of sign epistasis for the strong selection and weak mutation (SSWM) regime, where a non-recombining population evolves toward fitness peaks along accessible paths. In a smooth landscape (which possibly contains magnitude epistasis but no sign epistasis), all \textit{direct} paths to a peak 
are accessible, and these are the only accessible paths. Quantitatively, a genotype at Hamming distance $d$ from a peak has $d!$ direct paths, corresponding to substituting the $d$ beneficial mutations in any order, and all of these paths are accessible. When sign epistasis is present, the sign of fitness change produced by some of these mutations depends on the genetic background, and therefore not all the $d!$ direct paths are fitness-increasing \citep{weinreich2005perspective}. Sign epistasis therefore provides a mechanism by which genetic interactions can constrain evolution. These ideas were presented more formally in \cite{crona2013peaks} in the language of {\em fitness graphs}, which have now become a mainstay of the accessibility literature (see \cite{de2009exploring} for an earlier application).  

\subsubsection{Empirical fitness landscapes}

{Many of the theoretical developments that we review here have been motivated by the increasing availability of empirical fitness landscapes.}
The recent surge of interest in the experimental exploration of fitness landscapes was initiated by 
a seminal study by \cite{weinreich2006darwinian} on a combinatorially complete set of $2^5 = 32$ mutants in a 
$\beta$-lactamase gene in {\em E. coli} (see \cite{Ogbunugafor2026} for review). By constructing and phenotyping all intermediates separating a highly efficient 5-fold mutant enzyme from its low-efficiency ancestor, the authors showed that, out of the $5!=120$ direct paths from the ancestor to the full mutant, only 18 were selectively accessible, in the sense that resistance to the antibiotic (the natural fitness proxy in this system) increased in each step. Subsequent studies of small data sets of similar size have established the prevalence of sign epistasis in these landscapes  \citep[e.g.][]{de2009exploring,carneiro2010adaptive,kvitek2011reciprocal,franke2011evolutionary,salverda2011initial,schenk2013patterns,Maharjan2013,tufts2015epistasis}, and concluded that they generally display
a level of ruggedness that is intermediate between completely random and additive landscapes 
\citep{de2014empirical,szendro2013quantitative,aguilar2017thousand}. 

The past decade has seen a massive increase in the scale of fitness landscapes that can be 
surveyed experimentally \citep{Bank2016,Kinney2019,pokusaeva2019experimental,Nishikawa2021}, and 
current studies systematically evaluate fitness and related phenotypes for hundreds of thousands
of combinatorial mutants \citep{moulana2022compensatory,papkou2023rugged, westmann2024highly, westmann2026adaptive}. While this work has opened up new vistas to fitness landscape research and provides empirical grounding for many of the conceptual questions addressed in this review, 
the number of empirical fitness landscapes that can be characterized at this level of detail remains limited. This makes it difficult to draw general conclusions about the topography of natural fitness landscapes, and highlights the need for more systematic empirical measurements across diverse organisms and biological scales.

\subsubsection{Path complexity: the ``second question" of navigability}
\label{Sec:Pathcomplexity}

Based on our observations so far,  we can identify two different ways in which epistasis constrains landscape navigation. The first and most studied issue is that epistasis creates rugged (multi-peaked) landscapes, which means that a population may become stuck at a suboptimal peak. The topic of peak \textit{accessibility} in rugged landscapes is discussed in detail {in Sect.~\ref{Sec:PeakAccessibility}}. The second question, which has received less attention, is whether epistasis may slow down evolution by affecting the lengths of paths. A hypothesis, studied predominantly in the computational complexity literature \citep{johnson1988how,schaeffer1991local,matousek2006random,kaznatcheev2019computational,kaznatcheev2024local}, is that accessible paths to fitness peaks may be so long that they cannot be traversed within (computationally) feasible time scales. This problem is orthogonal to the more traditional concern that evolution stalls at local maxima, since it can apply even to single-peaked landscapes. For example, work by \cite{matousek2006random} on the performance of local search algorithms presented single-peaked landscapes on which uniform adaptive walks do not terminate in polynomial time \citep{kaznatcheev2019computational}; that is, the number of steps to the fitness peak from a random starting point grows faster than any power of $L$ (the number of loci) when $L$ is sufficiently large. In the same vein, \cite{kaznatcheev2019computational} constructed certain {\em semismooth} landscapes (defined as possibly having sign epistasis but not RSE) where the single fitness peak cannot be reached by a greedy walk in polynomial time. To drive home the implication, \cite{kaznatcheev2019computational} mentions that in one of the landscapes studied therein, a selective process ``with a genotype on just 120 loci, and, with a new set of point-mutants and selective sweep at a rate of one every second, would require more seconds than the time since the Big Bang" to arrive at the fitness peak. However, one should be careful not to draw conclusions about evolutionary processes in nature from this work. On the one hand, it is empirically unknown whether such landscapes exist or are common, and, from a biologist's perspective, the studied models, inspired by the goal to develop fast search algorithms in computer science, represent gross oversimplifications of the evolutionary process in nature (see also Section 4). Nonetheless, we note that the topic remains underexplored, and unlike the case of fitness peaks reviewed above, general bounds connecting the number and lengths of accessible paths to epistasis are sparse in the literature. We hope that future work clarifies the relevance of this ``path complexity" question in the navigation of fitness landscapes.

\subsection{Reaching high-fitness peaks  in rugged landscapes}
\label{Sec:PeakAccessibility}

Which factors determine the accessibility of high-fitness peaks and the frequency with which they can be reached by evolution?
Intuitively, one expects the accessibility of high-fitness peaks to correlate negatively with the number of peaks, and this expectation was confirmed by a comparative study of small empirical landscapes mentioned previously \citep{szendro2013quantitative}. However, a few larger empirical fitness landscapes have been reported to be both rugged and navigable \citep{papkou2023rugged, westmann2024highly}. Meanwhile, theoretical studies of probabilistic fitness landscape models with tunable degrees of ruggedness suggest that the relationship between ruggedness and accessibility is considerably more complex. Such studies 
can help identify general fitness landscape features that enable evolving populations to readily access high-fitness peaks. Empirically measured fitness landscapes can then be examined to determine whether they exhibit such features.

\subsubsection{Random and structured landscapes}

The most detailed understanding of peak accessibility has been achieved so far for a model where fitness values are assigned randomly to genotypes
as independent, identically distributed random variables \citep{kauffman1987towards}, also referred to as the \textit{House-of-Cards (HoC) model}. {HoC landscapes are frequently used  
as a highly rugged null model for the 
comparison to empirical data} \citep{oros2026}. In the HoC model, the probability of existence of at 
least one accessible path between two genotypes with fitness $f$ and $f' > f$ increases abruptly from near zero to near unity at a threshold value of the difference $F(f') - F(f)$, where $F(x) = \mathrm{Prob}[f \leq x]$ is the cumulative
fitness probability distribution \citep{SchmiegeltKrug2023}. This implies that 
high fitness peaks are likely to be accessible from distant low fitness genotypes, and as a consequence possess surprisingly large \textit{adaptive basins} \citep{pahujani2025complexity,oros2026}. 

When fitness correlations are introduced into the HoC model, the ruggedness of the landscapes generally decreases, but the effect on accessibility varies dramatically depending on the specific setup. For example, in the 
\textit{NK model} the tuning parameter 
is the number of epistatic interaction partners $K$ of a particular locus \citep{kauffman1989nk}. When $K$ is small compared to the total number $L$ of loci, the landscape has low ruggedness. Nevertheless, the probability of existence
of accessible paths between distant genotypes in the NK model tends to zero irrespective of the fitness difference when $L$ becomes large at constant $K$,
implying that these landscapes are much less accessible than their HoC counterpart
(which corresponds formally to $K=L-1$). The low accessibility is due to the sparseness of the network of epistatic interactions, which leads to 
the emergence of motifs of reciprocal sign epistasis that are independent of genetic background and therefore block all long-distance accessible paths traversing the genotype space \citep{hwang2018universality}. This shows that local fitness peaks do not necessarily constitute the main obstacles to landscape navigation.

The results for the NK model contrast sharply with the accessibility properties of the \textit{RMF model}, where the maximally rough HoC landscape is smoothened by an overall fitness gradient towards a reference sequence \citep{Neidhart_2014}. In this case, any degree of smoothening makes the reference genotype accessible with high probability \citep{hegarty2014existence}. Taken together, the examples of the NK and RMF model indicate that, by itself, landscape ruggedness is generally a poor predictor of peak accessibility, because the path structure induced by the fitness correlations differs markedly between the two models. Thus, the potential to navigate long distances across the landscape is not necessarily inhibited by ruggedness. 

\subsubsection{Global and universal epistasis}

A pervasive pattern in empirical data is the occurrence of \textit{global epistasis}, which we here define as epistasis arising from a nonlinear phenotype-fitness map applied to a non-epistatic phenotypic trait that is a linear function of the genotype
\citep{otwinowski2018inferring,johnson2023epistasis,diaz2023global}. When the phenotype-fitness map is non-monotonic, such a mapping can give rise to rugged, multipeaked landscapes \citep{rokyta2011epistasis,hwang2017genotypic}. Recent theoretical work has shown that this type of global epistasis imposes a constraint on the resulting
fitness landscape that has been termed universal epistasis \citep{crona2023geometry}, and that has important consequences for the accessibility of fitness peaks. 

To explain the concept of universal epistasis, it is useful to think of genotype sequences as sets of mutations, a representation that can be made precise for binary sequences \citep{das2020predictable}. Specifically, a binary sequence of length $L$ is mapped to the subset of loci with allele 1, signifying the presence of a mutation. For example, for $L=4$, the sequence (0110) maps to the set $\{2,3\}$. In the spirit of `beanbag genetics' \citep{haldane1964defense}, a genotype is viewed as a bag of mutations (relative to a reference sequence). To define universal epistasis, consider adding a set of mutations 
to two different background genotypes $g,g'$, where one (say $g'$) is a subset of the 
other. The mutations that are added are not contained in either of the background
genotypes. Then universal epistasis holds if the difference 
between the fitness effects of adding the same set of mutations to the two backgrounds has a definite sign, such that the compound effect of the mutations is always larger (positive epistasis) or smaller (negative epistasis) in background $g$ compared to $g' \subset g$ \citep{crona2023geometry}. 

When universal epistasis is negative, it can be proved that any peak genotype is accessible via all direct paths from all its subset and superset genotypes \citep{krug2024evolutionary,pahujani2025complexity}. In the example above, if genotype $(0110)$ is a peak, it is accessible from $(0000),(0100),(0010),(1110),(0111)$ and $(1111)$. This implies a lower bound on the size of adaptive basins of the peaks, as well as the existence of (at least) two genotypes from which \textit{all} peaks can be accessed: The ``wild-type" $(00\dots0)$ and the ``full mutant" $(11\dots1)$. In this sense, universal negative epistasis is associated with the ease of evolutionary navigation to fitness peaks. Negative universal epistasis can be viewed as a generalization of 
diminishing returns epistasis, which is a common feature of biological systems and has been frequently observed in evolution experiments \citep{chou2011diminishing,schoustra2016diminishing,johnson2023epistasis}. 

\subsubsection{Alphabet cardinality and allele graph connectivity}
Real fitness landscapes are hyper-astronomically large \citep{louis2016contingency} and any landscape that has been empirically measured or theoretically modeled is only a subset of the larger landscape. Sub-landscapes can arise when the landscape is generated from a limited number of loci or when the number of possible alleles at each locus (also known as the \textit{alphabet cardinality}) is limited. This can result in an underestimation of the number of accessible paths in a landscape, since allowing for more alleles at each locus can open up extra-dimensional bypasses to the global peak \citep{WuEtAl2016, ZagorskiEtAl2016, SrivastavaRozhonovaPayne2023}. Moreover, \cite{SchmiegeltKrug2023} showed that the threshold value of the fitness difference between the initial and final genotype in the HoC landscape ($F(f') - F(f)$, see section above), required for a high probability of an accessible path between the two genotypes, decreases with the number of alleles at each locus. 

On the other hand, \cite{SchmiegeltKrug2023} demonstrated that decreasing the connectivity between alleles in the \textit{allele graph} raises 
{this threshold}, 
thereby reducing the accessibility of high-fitness peaks. This reduced connectivity of alleles becomes biologically relevant when examining protein fitness landscapes, wherein certain amino-acid transitions may not be possible via single-nucleotide changes due to the constraints imposed by the genetic code \citep{RozhonovaEtAl2024}. Therefore, failing to account for the genetic code may lead to an overestimation of the global peak \textit{reachability} in protein fitness landscapes. A similar reduction in the overall landscape \textit{navigability} due to low allele-graph connectivity was also observed in the Biomorphs fitness landscape, which is a landscape of two-dimensional developmental phenotypes \citep{martin2024bias}, wherein alleles can only change by a single unit per mutational step \citep{srivastava2024adaptive}.

Merely having more accessible paths to the global peak does not guarantee that the global peak will be reached by evolving populations: Studying RMF landscapes with varying levels of ruggedness, \citet{SrivastavaRozhonovaPayne2023} showed that even though the number of accessible paths to the global peak grew with increasing alphabet cardinality, the probability of reaching the global peak (as measured by the fraction of accessible paths to the global peak, out of all accessible paths starting in the antipodal genotypes of the global peak) decreased.

\subsubsection{{Basin size and reachability}}
The concern expressed by \cite{wright1932roles} that rugged landscapes may impede evolution naturally suggests the intuition that peaks in highly rugged landscapes have small \textit{adaptive basins}. 
In the light of recent work, however, this intuition needs to be revisited (see \citet{hernando2019anatomy} for a related discussion
in the context of combinatorial optimization). 

For an empirically-motivated landscape model of bacterial fitness in {the presence of} antibiotics, \citet{das2020predictable} found that peaks have large overlapping adaptive basins. \cite{papkou2023rugged} found an empirical landscape of the metabolic enzyme dihydrofolate reductase (DHFR) to be highly rugged, and all high-fitness peaks essentially shared a single and very large adaptive basin. Similarly, the landscape of the bacterial transcription factor TetR was found to be highly rugged yet also highly navigable \citep{westmann2024highly}. 
Further, as mentioned above, the highly rugged HoC model also has large adaptive basins; for an alphabet cardinality of $4$, an average peak basin contains about 52.8$\%$ of all genotypes when $L$ is large \citep{oros2026}. These findings challenge the idea that high ruggedness necessarily restricts accessible paths. However, it is not yet known what underlying epistatic patterns can make a landscape rugged while generating large adaptive basins, and whether such landscapes are common in nature.

Another related concern about adaptive basins is their utility, in general, for predicting the \textit{reachability} of peaks. The study of the DHFR landscape by \cite{papkou2023rugged} found a high correlation between peak fitness and basin sizes, and suggested this as an explanation for the high reachability of high-fitness peaks.  
A similar positive relationship between peak heights and basin sizes in both RMF landscapes and empirical landscapes of the bacterial ParD–ParE toxin–antitoxin system was reported by \citet{SrivastavaRozhonovaPayne2023}.
However, a numerical study of several theoretical and empirical landscapes \citep{li2025probability} concluded that a positive correlation between peak height and basin size is general, and it does not in itself imply high reachability
(see also \citealp{oros2026}). The reachability instead depends on specifics of the landscape \citep{li2025evaluation,hunter2026rmf}. Moreover, in two empirical fitness landscapes studied by \cite{reia2020analysis}, reachability of peaks starting from the wild-type sequence tended to increase with the number of accessible paths, but the relationship between the two was quite irregular.
Thus, adaptive basin sizes or the number of accessible paths may not have sufficient information about the reachability of peaks by evolving populations in general. This is a serious limitation and thus these quantities should be regarded with caution as measures of navigability. More effective measures may rely on further details of how basins or paths are distributed across genotype space.

\subsection{Genotype-phenotype maps and neutral component graphs}

So far, we have focused on genotype–fitness maps that directly assign fitness values to genotypes by bypassing intermediate levels of biological organization and invoking a somewhat abstract concept of fitness. However, biological systems are organized across multiple concrete scales. Information is stored in DNA or RNA sequences, where molecular phenotypes such as RNA secondary structure are relevant; these sequences are transcribed and translated into amino acid chains, where protein structure and binding affinity become the pertinent phenotypes. Multiple cells assemble into tissues, which in turn constitute an organism, and at these higher levels, more macroscopic traits, such as tissue morphology or body plan, are the fitness-determining phenotypes. Thus, genotypes can map to different phenotypes depending on the biological scale and question under consideration \citep{Pigliucci2010GP, Gjuvsland2013BridgingGP}. These \textit{genotype–phenotype (GP) maps} characterize how mutations generate phenotypic variation, for instance, by quantifying how often mutations preserve the phenotype or how many distinct phenotypes can be generated via single mutational steps.

\subsubsection{Properties of GP maps}

Despite the variation in scale and in the definition of fitness-relevant phenotypes, the literature on genotype–phenotype (GP) maps, which has been shaped strongly by empirical findings and extensive computational modeling, has identified several ``universal properties" of these maps \citep{Ahnert_2017} which could arise due to the existence of unconstrained loci in the genotype that do not affect the phenotype \citep{Greenbury2015, Weiss2018}. Multiple reviews exist on these shared properties of GP maps (see, e.g., \citet{Manrubia_2021}), but for completeness, we briefly mention them here. GP maps show (i) Genotypic redundancy, i.e., there are many more genotypes than phenotypes. (ii) Phenotypic bias, i.e., the distribution of the number of genotypes per phenotype is skewed and Zipfian in character, with a few phenotypes realized by many genotypes, whereas most phenotypes are represented by only a small number of genotypes. (iii) High \textit{robustness} of phenotypes, wherein the robustness scales logarithmically with the frequency of the phenotype $\rho_p \sim a+ b  \cdot \log (f_p)$ rather than linearly. (iv) A positive correlation between phenotypic robustness and \textit{phenotypic evolvability}. (v) Shape-space covering property, wherein the majority of phenotypes lie in close proximity to any genotype. (vi) Exponential bias towards ``simple" phenotypes, i.e., phenotypes that require less information to be encoded algorithmically \citep{Johnston2022SymmetrySimplicity}.

These GP map properties play a crucial role in shaping evolutionary outcomes \citep{Cowperthwaite2008, SchaperLouis2014ArrivalFrequent, Dingle2015}. For instance, \cite{SchaperLouis2014ArrivalFrequent} studied how GP map properties shape the evolution of haploid, Wright-Fisher populations exploring a neutral plateau and found that more frequent phenotypes are discovered orders of magnitude faster than less frequent phenotypes. The authors also used a 12-locus RNA GP map, initialized populations on a large neutral network connected by mutations to two fitter target phenotypes with different probabilities of being reached, and showed that more frequent phenotypes can fix even when less frequent but fitter alternatives are accessible. Further, \cite{Cowperthwaite2008} showed that naturally occurring functional RNA molecules are biased towards abundant phenotypes. These results highlight the importance of the GP map in addition to the fitness values of phenotypes in steering evolutionary dynamics. This phenomenon has been termed, ``ascent of the abundant" \citep{Cowperthwaite2008} or the ``arrival of the frequent" \citep{SchaperLouis2014ArrivalFrequent}. Put differently, this shows how large neutral networks in GP maps can constrain adaptation by preventing the population from reaching high fitness peaks \citep{Cowperthwaite2008}. Interestingly, \cite{Dingle2015} found that randomly sampled genotypes in the RNA GP map predicted the distribution of properties of naturally occurring non-coding RNA accurately, suggesting that abundant, findable phenotypes may still have a reasonably high fitness, since small structural differences may not be very consequential for function.  

In this vein, arguments from algorithmic information theory provide a non-adaptationist explanation for the occurrence of simple, symmetric structures: because such structures require less information to encode, they are more likely to arise as phenotypic variants through random mutation \citep{Johnston2022SymmetrySimplicity}. More recently, \cite{MaloneEtAl2025} used a computational model of leaf development to show that developmental bias towards simpler leaf morphologies could explain phylogenetic trends even in the absence of selection.  

\subsubsection{Evolvability in NC graphs}

Whereas most of the above studies examined GP maps without selection, \citet{Greenbury2022NavigableGP} explicitly linked the GP map structure to the \textit{navigability} of fitness landscapes. In a given GP map, it is not obvious which phenotype corresponds to the highest fitness, and this can vary with the environment or other extrinsic factors. Thus, when studying adaptation on GP maps, a \textit{target} phenotype is sometimes arbitrarily chosen and assigned the highest fitness. The remaining phenotypes are then either assigned random fitness values or fitness values based on similarity to the target phenotype. \citet{Greenbury2022NavigableGP} randomly assigned fitness values to phenotypes for three biologically realistic GP maps and randomly chose target phenotypes that were always assigned the highest fitness. They found that even under this random, uncorrelated fitness assignment, the GP map properties mentioned above increased the accessibility of the target phenotypes. This implies that navigability of fitness landscapes is facilitated by the features shared by most GP maps.

Further, \citet{Srivastava2026} adopted a coarse-grained view of the GP map by studying it as a graph of \textit{neutral components (NCs)}, wherein the NCs constituted the nodes of the NC graph, whereas edges were defined by the presence of point mutations connecting two NCs. Using this, the authors identified the \textit{NC evolvability} ($\epsilon_{NC}$) as the key GP map property that determines the number of peaks and the landscape navigability under random fitness assignment. Note that the NC evolvability would be equivalent to the phenotypic evolvability if all genotypes mapping to a given phenotype were connected within a single neutral component; however, they can sometimes be fragmented into multiple disconnected components. The authors found the probability of an NC being a fitness peak to be inversely proportional to its evolvability, i.e., $P_{NC}(peak) = \frac{1}{\epsilon_{NC} +1}$. This relation generalizes a classical result for the genotypic HoC landscape \citep{kauffman1987towards} to GP-map-based landscapes, see Sect.~\ref{Sec:Synthesis} for further discussion. Moreover, the average fitness of peaks was found to increase with their evolvability, regardless of the distribution from which the fitness values were drawn. This implies that when evolvability scales positively with NC size, larger peaks also have higher fitness. Finally, the authors derived a minimum threshold of mean evolvability of a GP map above which the corresponding random fitness landscapes have non-zero navigability.

However, as mentioned in the sections above, the navigability of a landscape only guarantees the existence of \textit{an} accessible path to the target, but that path may not be traversed by evolving populations. This is especially true when populations become trapped on large neutral plateaus, which might take a long time to traverse \citep{ManrubiaCuesta2015}. To exclude enormously long accessible paths from contributing to the navigability of a target, \citet{Greenbury2022NavigableGP} set a computational threshold on the path length beyond which the search for an accessible path to the target was terminated. Further, \citet{Greenbury2022NavigableGP} defined a more dynamic measure called \textit{evolutionary navigability}, as the average probability that the adaptive walk
reaches a target phenotype from a source phenotype via an accessible path. Although \citet{Greenbury2022NavigableGP} showed that accessible paths between pairs of functional RNA sequences were also likely to be utilized under evolutionary dynamics (i.e., \textit{landscape navigability} and \textit{evolutionary navigability} were similar), in the case of the Biomorphs GP map, the \textit{landscape navigability} was high under a correlated fitness assignment, but a very small fraction of the evolving populations could reach the target phenotypes \citep{srivastava2024adaptive}. This could be because of the reduced allele graph connectivity and bias towards simple phenotypes in the Biomorphs GP map. Therefore, much like genotype-fitness maps, summary statistics for GP maps, such as landscape navigability, are useful, but can sometimes be incomplete predictors of the evolutionary dynamics on the resulting fitness landscapes.

\subsection{Synthesis of genotype-fitness map and GP map approaches: current state and future directions}
\label{Sec:Synthesis}
The accessibility literature has largely focused on complex genotype–fitness maps, but with a limited role for neutrality; conversely, the literature on empirically motivated GP maps has primarily emphasized GP map structure and typically focused on neutral networks or considered very simple fitness assignments, such as random phenotype-to-fitness maps. However, ultimately, GP maps and genotype–fitness maps are connected and represent different aspects of the study of evolution. There is therefore a need for closer integration of these two approaches.

In random genotype--fitness (GF) maps, alphabet cardinality, i.e., the number of alleles per site and the mutational connectivity between those alleles, plays a central role in determining landscape ruggedness, as it directly controls the number of mutational neighbors available to each genotype. For instance, in a HoC genotype-fitness map of genotype length $L$, with a fully mutationally connected allele graph with $a$ alleles,  each genotype has $(a-1) \cdot L$ neighbors. Thus, the probability of a random genotype being a fitness peak is $\frac{1}{(a-1)\cdot L+1}$ \citep{kauffman1987towards}. Further, the average peak height when the fitness values are drawn from a uniform distribution between 0 and 1 is $$ \langle f_{GF}^{\text{random}} \rangle = \frac{(a-1)\cdot L+1}{(a-1)\cdot L+2}.$$ On the other hand, in GP maps with random fitness assignment, ruggedness is governed by the evolvability of neutral components i.e., $\epsilon_{NC}$, which 
equals the number of distinct phenotypes accessible from a given neutral network. Each neutral component (NC) has $\epsilon_{NC}$ neighbors and thus the probability of a random NC being a fitness peak is $\frac{1}{\epsilon_{NC}+1}$ and the average peak height with uniformly distributed fitness values is $$\langle f_{GP}^{\text{random}} \rangle =  \frac{\epsilon_{NC}+1}{\epsilon_{NC}+2}.$$  
Therefore, it is evident that alphabet cardinality ($a$) and NC evolvability ($\epsilon_{NC}$) are conceptually similar, as both correspond to measures of node degree in genotype space and in the NC graph, respectively, and determine the average number of peaks and the average peak height in random GF and GP maps. Importantly, the alphabet cardinality and allele graph connectivity can also influence the mean evolvability of a GP map. 

However, these quantities differ in an important way: alphabet cardinality and allele graph connectivity are typically uniform or vary very slightly across all genotypes (e.g., via the codon table), at least when the genotype graph is complete (substantial inhomogeneity in genotype connectivity can arise due to missing genotypes, as is often the case in large-scale empirical data sets, see \citet{papkou2023rugged,westmann2024highly,westmann2026adaptive,li2025probability}). In contrast, NC evolvability can vary over several orders of magnitude within a GP map due to phenotypic bias \citep{schaper2012epistasis}. Furthermore, increasing alphabet cardinality and allele graph connectivity necessarily expand the total size of the genotype space, thereby uniformly increasing mutational connectivity. By contrast, evolvability depends on the size and structure of neutral components and is therefore determined by the genotype–phenotype mapping itself; as a result, the same number of genotypes can give rise to markedly different distributions of evolvability \citep{Greenbury_2016}.

As discussed in the sections above, increasing alphabet cardinality, allele graph connectivity, and the average NC evolvability all enhance the navigability of the corresponding fitness landscapes. However, none of these quantities guarantees that evolving populations will reach a target genotype/phenotype (i.e., a fitness peak). Instead, populations may become trapped either at local optima or in large neutral networks called entropic traps, with the latter having been noted to be easier to escape \citep{vanNimwegenCrutchfield1999}. For example, \citet{srivastava2024adaptive} showed that in the Biomorphs fitness landscape, on average, 32\% of adaptive walks terminated in an entropic trap (defined as walks that continued beyond the threshold of $10^6$ steps), whereas 68\% of the walks terminated on a local peak. In addition, mutation bias \citep{cano2020mutation, cano2022mutation} and phenotypic bias \citep{SchaperLouis2014ArrivalFrequent, dingle2022phenotype, martin2024bias} can also steer populations away from the global peak despite the existence of accessible paths. 

Therefore, landscape navigability should not merely be quantified by epistasis, number of peaks, accessible paths to the global peak, or the basin sizes of the genotype-fitness map. Rather, there is a need for measures that focus on how these structural properties influence dynamical outcomes, thereby implicitly capturing their effects.

To this end, we propose a new measure of navigability, the \textit{navigational ruggedness} $\mathcal{R_{N}}$:

$$\mathcal{R_{N}} = 1-\sum_{\text{peak} \in \{ \text{all peaks}\}}P_{\text{reach}}(\text{peak}) \cdot f_{\text{rel}}(\text{peak}),$$
where the summation is over all peaks in the landscape, $P_{\text{reach}}(\text{peak})$ is the probability of reaching a given peak under a specified evolutionary dynamics that terminates at peaks, and $f_{\text{rel}}(\text{peak})$ is the relative fitness of the peak with respect to the global peak fitness. If high peaks have a high probability of being reached, then 
$\mathcal{R}_N \ll 1$. 

\begin{figure*}[h!]
\centering 
\includegraphics[width=\linewidth]{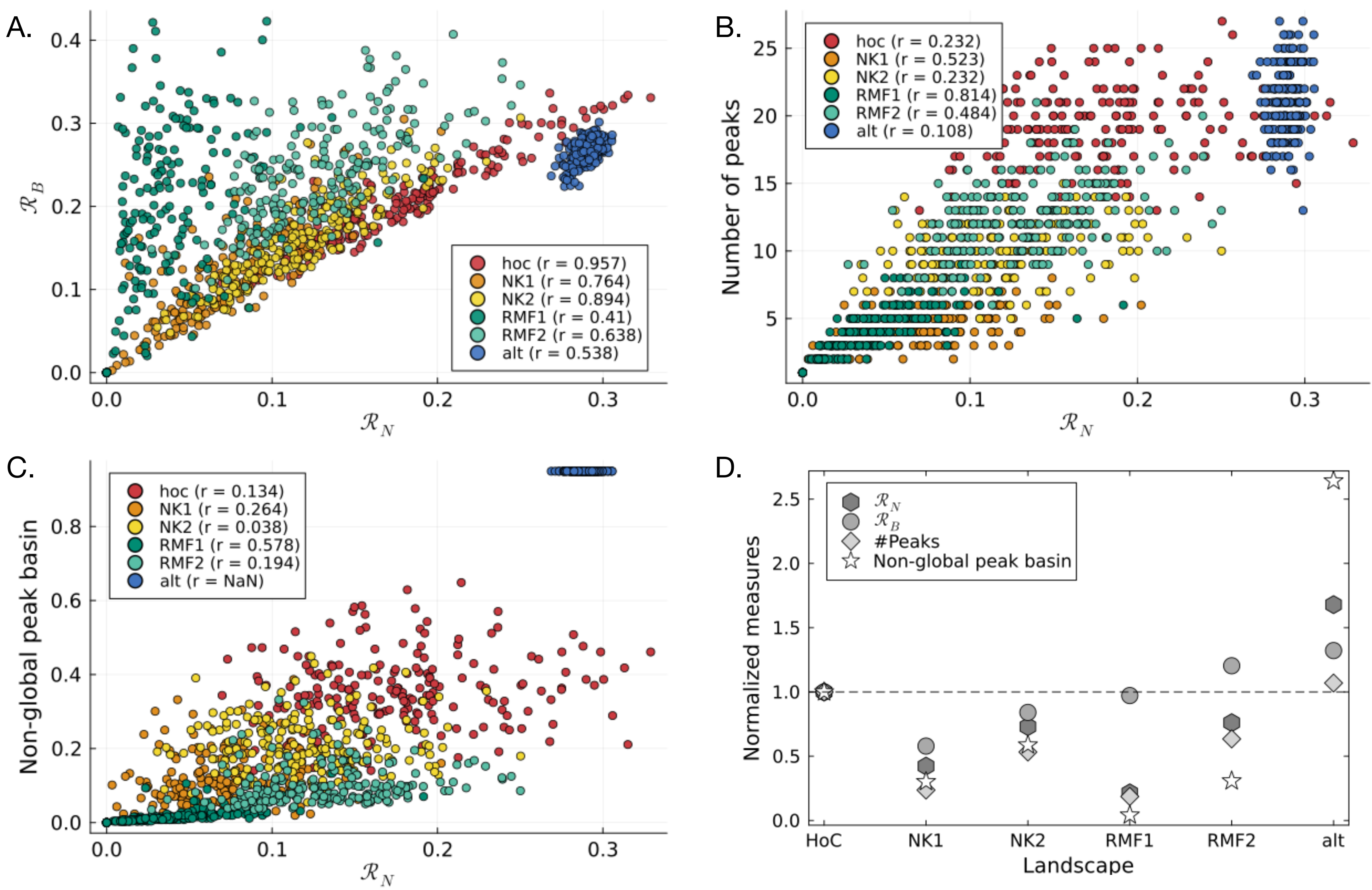} 
\caption{\textbf{Comparison of ruggedness measures across fitness landscapes.} \textbf{A.} Baseline ruggedness versus navigational ruggedness. \textbf{B.} Number of fitness peaks versus navigational ruggedness. \textbf{C.} Relative size of the non-global peak basins versus navigational ruggedness. Each point corresponds to one of 200 landscapes from a given class (indicated by different colours; Pearson correlation coefficients between measures are shown in the figure legends). \textbf{D.} Ruggedness measures averaged over 1000 landscapes of each class and normalized by their corresponding values for a House-of-Cards (HoC) landscape. Measures shown are navigational ruggedness (hexagons; estimated from 5000 random adaptive walks per landscape), baseline ruggedness (circles), number of peaks (diamonds), and the number of genotypes outside the global-peak basin (stars). All landscapes have genotype length $L=4$ and alphabet cardinality $a=4$. Landscape classes include a House-of-Cards landscape (HoC); two NK landscapes (NK1: $N=L$, $K=1$, intermediate epistasis; NK2: $N=L$, $K=2$, high epistasis, with interactions between adjacent loci and periodic boundary conditions); two Rough Mount Fuji landscapes (RMF1 and RMF2, with intermediate and high ruggedness, respectively); and a custom-designed alternating-fitness landscape (alt; see Methods). Unless otherwise stated, random fitness contributions were drawn independently from a normal distribution, $\mathcal{N}(0,1)$.
}\label{fig:nr}
\end{figure*}
In Figure \ref{fig:nr}, we show four different measures of ruggedness: (i) navigational ruggedness $\mathcal{R}_N$, (ii) baseline ruggedness, defined using the equally weighted average peak fitness, i.e. 
$$\mathcal{R}_B = 1-\sum_{\text{peak} \in \{  \text{all peaks}\}} \frac{1}{N_{\text{peaks}}} f_{\text{rel}}(\text{peak}),$$ 
where $N_{\text{peaks}}$ is the total number of peaks in the landscape, (iii) the number of genotypes that are peaks, and (iv) the relative number of genotypes that do not belong to the adaptive basin of the global peak. We computed these measures for the HoC, NK ($K=1$ and $K=2$ ), RMF  ($c=1$ and $c=2$; see Methods)
landscapes, and a fitness alternating landscape which was designed to be highly rugged, with many low fitness peaks with large basin sizes. All landscapes had $L=4$ loci and alphabet cardinality $a=4$. To enable comparison, we plotted the remaining three measures against the navigational ruggedness in Figures \ref{fig:nr} A, B and C and the average value of all the measures over 1000 landscapes relative to the values for the HoC landscape in Figure \ref{fig:nr} D.

Whereas the baseline ruggedness $\mathcal{R}_B$ is based solely on the mean peak fitnesses, the navigational ruggedness $\mathcal{R}_N$ also accounts for the probability of reaching the peaks. We find that $\mathcal{R}_B \geq \mathcal{R}_N$ for the HoC, NK, and RMF landscapes but not for the alternating landscape, due to the anti-correlation between the peak heights and basin sizes in the alternating landscape. Further, the two measures are correlated to varying degrees; while HoC landscapes have a very high correlation ($r \approx 0.95$), RMF landscapes have a much lower correlation  ($ 0.4<r<0.65$) (see Figures \ref{fig:nr} A and D). 
Next, the number of peaks and the relative basin size of all non-global peaks are weakly correlated with the navigational ruggedness for most landscape classes (see Figures \ref{fig:nr} B and C).Note that for the alternating landscape class, the relative basin size of all non-global peaks is identical across all 200 realizations; since this quantity has zero variance, its Pearson correlation with $\mathcal{R}_N$ is undefined (NaN) in Figure~\ref{fig:nr} C. Put together, these results show that no one ruggedness measure is truly indicative of the ease of landscape navigation.

Notably, the navigational ruggedness $\mathcal{R}_N$ is a summary of various landscape features, such as peak heights, basin sizes, etc., which are all relevant for navigability. While the measure does not inform us about the specific features facilitating landscape navigation, any causative feature should correlate with the navigational ruggedness and can thus be identified. For instance, it can be seen in Figures \ref{fig:nr} A and D, that the baseline ruggedness $\mathcal{R}_B$ is not a good indicator of the navigational ruggedness $\mathcal{R}_N$ for the RMF and alternating landscapes: in the former, it overestimates the navigational ruggedness, whereas it underestimates it in the latter. Conversely, the relative basin size of all non-global peaks underestimates the navigational ruggedness in RMF landscapes and overestimates it in the alternating landscape (see Figures \ref{fig:nr} C and D). Moreover, navigational ruggedness $\mathcal{R}_N$ of RMF landscapes remains substantially lower than the baseline ruggedness $\mathcal{R}_B$ and also much lower than the navigational ruggedness $\mathcal{R}_N$ of NK and HoC landscapes, reinforcing the point that the global peak is readily reachable in RMF landscapes \citep{hegarty2014existence}. This is also apparent from the low relative basin size of the non-global peak genotypes. Thus, the navigational ruggedness $\mathcal{R}_N$ captures important features of landscape navigation that are not captured by the other measures. 

In Supplementary Figure \ref{sfig:2}, we examine how navigational ruggedness depends on landscape dimensionality, quantified by the alphabet cardinality, $a$. We find that the effect of increasing dimensionality is largely determined by its impact on baseline ruggedness. Consequently, because baseline ruggedness is sensitive to the distribution of fitness values used to generate the landscape, navigational ruggedness exhibits a similar dependence. For HoC landscapes generated by drawing fitness values from a normal distribution and subsequently normalizing them to the interval $[0,1]$, navigational ruggedness increases with increasing dimensionality. In contrast, for HoC landscapes generated from a uniform fitness distribution, navigational ruggedness decreases as dimensionality increases. As a comparison, we also examined smooth RMF landscapes and found that navigational ruggedness increases with dimensionality, consistent with the trend reported in \citet{SrivastavaRozhonovaPayne2023}.


\section{Perspectives on fitness landscape navigation beyond adaptive walks and static landscapes}
Up to here, our review has focused on studies that quantify fitness landscape navigation in the context of strong selection and weak mutation and thus implicitly or explicitly assume that adaptation occurs as discrete, monomorphic, fitness-increasing steps in the genotype space. Moreover, these studies assume that the mapping of genotype to phenotype or fitness is constant over time. Here, we highlight studies that have relaxed these assumptions and indicate future research directions in the context of navigability. Although it offers greater biological realism, this work is often limited to numerical approaches because the navigation process is difficult to quantify analytically.   


\subsection{Weak selection and clonal interference}
One key assumption that has facilitated the analytical study of adaptation on fitness landscapes is that of strong selection and weak mutation \citep{gillespie1983some}, where adaptation occurs through the sequential occurrence and fixation of beneficial mutations. However, natural populations often violate this assumption.  

A body of work examines the weak-mutation (sequential origin-fixation) regime without assuming strong selection, allowing deleterious mutations a nonzero fixation probability \citep{mccandlish2014modeling}. Here, evolution proceeds through walks that include fitness-decreasing steps, which therefore never terminate. Instead, the system converges to an equilibrium distribution over genotypes. This distribution depends only on fitness values and not on the landscape neighborhood structure when the underlying Markov chain is reversible \citep{berg2004adaptive,sella2005the,manhart2012}.  \citet{mccandlish2014inevitability} showed that in the HoC model, the expected selection coefficient of the first substitution starting from equilibrium is negative. This is because equilibrium populations spend most of their time at rare, high-fitness genotypes, making an initial deleterious step likely. Relatedly, \citet{mccandlish2013findability}  found that the mean time to reach a focal genotype starting from equilibrium is longer when the genotype's fitness is lower, but bears no simple relationship to fitness rank, indicating that the shape of the landscape plays a crucial role.
\cite{servajean2023impact} found that the mean fitness of the first peak reached in a landscape starting from a random genotype was often maximized at an intermediate population size rather than in the strong-selection limit of infinite population sizes. Related to Wright's Shifting Balance theory \citep{Wright1982,coyne1997perspective,joshi1999shifting}, \cite{servajean2023impact} noted that larger population sizes bias evolution toward larger fitness-increasing steps but reduce exploration of the fitness landscape through deleterious or weakly beneficial mutations, indicating a trade-off relevant to efficient navigation of landscapes. 

The weak-mutation regime allows for drift but does not include cases in which multiple genotypes can segregate simultaneously leading to potential interference between beneficial variants and hitchhiking of deleterious variants. 
When the assumption of sequential origin-fixation is relaxed, a population can escape a local fitness peak without fixing a deleterious mutant through a process called `stochastic tunneling' \citep{iwasa2004stochastic,weinreich2005rapid}. In this process, the deleterious single mutants produced from the peak genotype persist in the population at low frequency and eventually produce a double mutant that is fitter than the peak genotype, which can become fixed in the population. For a given mutation rate, the critical population size above which the tunneling process dominates over sequential fixation was quantified by \cite{weinreich2005rapid}.  \cite{weissman2009rate} found that such a threshold population size exists also for wider fitness valleys comprising multiple mutations. Thus, again the population size was identified as an important factor influencing the navigation of the fitness landscape. 


Relatively few studies address clonal interference in the context of navigating larger rugged landscapes. \citet{szendro2013predictability} studied evolution on an experimentally-determined 8-locus {\em Aspergillus niger} landscape and an RMF variant, finding that increasing population size made dynamics `greedier' and more deterministic, as mutations of large effect were preferentially fixed (a similar behavior occurred at high mutation supply in an antibiotic resistance model by \citealp{das2025epistasis}). At even higher population sizes, second-step mutants capable of crossing fitness valleys arose more often, making dynamics more stochastic again due to the large number of such mutants. \citet{guo2019stochastic}, studying a landscape with many loci and sparse second-order epistasis, found that stochastic tunneling caused the system to `hop' between local peaks, with hopping intervals increasing over time, producing a logarithmic fitness increase reminiscent of slow energy relaxation in spin-glass physics.

Overall, these studies indicate that relaxing the strong-selection weak-mutation assumption can alter landscape navigation in complex, system-dependent ways, though systematic work on this question is still limited. Further, how drift and tunneling affect the ``path complexity'' problem from {Sect.}~\ref{Sec:Pathcomplexity} remains an open question.

\subsection{Recombination and genome organization}
The fitness landscape theory discussed so far considers the genotype space to be haploid and reproduction to be clonal. This is a gross simplification in a biological world in which recombination occurs both via sex and horizontal gene transfer. With recombination, the physical organization of loci in the genome matters, and recombination significantly affects the navigation of fitness landscapes by shuffling alleles or allele complexes between segregating genotypes \citep{bank2022epistasis}.

Theory predicts opposing effects of recombination on the speed of adaptation on rugged fitness landscapes operating at different time scales \citep{de2009exploring,nowak2014multidimensional}. In the short term, the adapting population 
traverses a relatively smooth portion of the landscape approaching a nearby local fitness peak. This process is facilitated by recombination through the combination of beneficial alleles arising in different backgrounds, a mechanism known as the Fisher-Muller effect
\citep{felsenstein1974evolutionary,kim2005adaptation,park2013rate}. Once a fitness peak has been reached, recombination with the majority peak genotype breaks 
down escape variants and traps the population at the peak
\citep{weinreich2005rapid,park2011bistability}.
In fitness landscapes with lethal genotypes, recombination concentrates the population
in regions where the viable network is highly connected, which {can lead} to a competition
between mutational robustness and fitness \citep{klug2019recombination}. 

\citet{li2024rapid} examined the cumulative effect of recombination and standing genetic variation on haploid adaptation (without new mutation) on tunably rugged landscapes, reinforcing the finding that high recombination hinders adaptation on rugged landscapes by breaking apart high-fitness combinations of alleles. The authors also found that adaptation from standing genetic variation with recombination followed a smoother-than-average path of allele fixations on the fitness landscape, indicating altered fitness landscape navigation. 

In the presence of recombination, genome organization (i.e., the physical location
of loci on chromosomes) becomes relevant for fitness landscape navigation. Recombination can strongly speed up adaptation in rugged fitness landscapes if the landscape displays a modular structure that is respected by recombination \citep{watson2011genome}. On the other hand, structural variants, such as chromosomal inversions, can suppress recombination and promote local adaptation \citep{kirkpatrick2006chromosome}; also translocations, fusions, or fissions can alter the recombination distance between interacting loci. Finally, ploidy determines the number of copies of each chromosome that an individual carries. Varying ploidy alters the topology of the fitness landscape (somewhat similar to altering alphabet cardinality) and generates the possibility for dominance or epistasis at a single locus (reviewed in \citealp{li2024dominance}). So far, little is known about the role of structural variants and ploidy on fitness landscape navigation.

\subsection{Evolution of the genotype space}
In general, fitness landscape theory assumes a landscape of a fixed size. The study of extra-dimensional bypasses can conceptually be seen as new paths revealed on an existing, larger, landscape. However, genome size evolution (for example via gene or genome duplications, gene loss, insertions, or deletions) could fundamentally change the genotype space in which adaptation occurs. In this vein, \cite{Martin2021} examined the architecture of the RNA sequence–structure GP map with indels and showed that the robustness to point mutations was correlated with robustness to insertions and deletions. The authors also showed that point mutations and indels have similar probabilities of generating alternative phenotypes. However, the possibility that the genotype space itself evolves has not received much attention in the field of fitness landscape theory, although the organization and size of genomes have been shaped by billions of years of evolution through processes such as duplications, deletions, and large-scale rearrangements \citep{hogeweg2012toward,Manrubia_2021}. It is unknown whether features such as the navigability or ruggedness of fitness landscapes are under selection over long time scales, and future work may uncover conserved structural features that facilitate the navigation of fitness landscapes.

\subsection{Environment, ecology and changing landscapes}
The study of fitness landscapes is deeply rooted in evolutionary biology, with a focus on mutation and selection as the main evolutionary forces driving landscape navigation. Moreover, the assumption of a fixed fitness topography implies constant selection pressures on evolutionary timescales, thereby implying that fast environmental fluctuations and frequency-dependent interactions can be ignored. However, recent empirical studies have called into question the common separation of ecological and evolutionary timescales, and the study of fitness landscapes is increasingly considering ecological factors in various ways.

Firstly, abiotic environmental change, e.g., via seasonality, catastrophic events, climate change, etc., can affect navigation of fitness landscapes when the topography of the landscape is altered, and increasing empirical evidence suggests that fitness landscapes vary greatly across environments (reviewed in \citealp{bank2022epistasis}). \cite{mustonen2009fitness} coined the word ``fitness seascapes" for fitness landscape theory that incorporates the effect of fluctuating environments on evolution. Regarding navigation on fitness seascapes and other environment-dependent fitness landscapes, some notable issues are the potentials for long-term maintenance of polymorphism (see \cite{bank2022epistasis} for a discussion) and irreversibility of adaptive walks \citep{das2022driven}. A significant application has been to antimicrobial resistance \citep{mira2015adaptive,ogbunugafor2016adaptive,das2020predictable,king2025fitness}, and recent work by \cite{manivannan2026deconstructing} examined how fitness seascapes vary across systems and scales, with particular focus on antimicrobial resistance. 

Environmental change can also lead to phenotypic plasticity. \cite{AncelFontana2000} showed that phenotypes produced by environmental perturbations are also mutationally accessible, making plastic responses a potential predictor of evolutionary trajectories. \cite{GarciaGalindo2023} found that the non-deterministic RNA GP that maps genotypes to the Boltzmann distribution over all possible secondary structures has the same structural properties as the deterministic RNA GP map.
Further, while \cite{Ancel2000} showed that the effect of plasticity on evolution is context-dependent, \cite{EspinosaSoto2011} used gene regulatory networks to show how plasticity facilitates navigation of genotype space toward new phenotypes. \citet{mall2026epigenetic} suggested that epigenetic switching can promote better navigation of rugged fitness landscapes by effectively smoothing them. Finally, a computational study of adaptation on RMF fitness seascapes found that the temporal variation can help recombining populations to escape from local fitness peaks and induce a long-term evolutionary advantage of recombination \citep{nowak2014multidimensional}.

Secondly, species interactions can generate a meta-fitness landscape of a community, in which the interacting species experience different and potentially opposing fitness peaks. One incorporation of species interactions was provided by \cite{kauffman1991coevolution}, who extended the probabilistic $NK$ fitness landscape model for ecological communities. When comparing adaptive walks on small fitness landscapes of two interacting species with adaptive walks on seascapes of the same dimension, \cite{hablutzel2026repeatability} found that species interactions led to higher endpoint repeatability, which may imply higher navigability of coevolutionary fitness landscapes. More research is needed to quantify the implications of species interactions for navigating larger fitness landscapes.

Thirdly, intraspecific interactions (e.g., via resource competition) could generate fitness seascapes that incorporate frequency-dependent interactions. \cite{amado2024ecological} simulated evolution on eco-evolutionary fitness landscapes that mapped genotype to resource traits, which mapped to fitness via a MacArthur-like resource-consumer model. Adaptation on the resulting fitness landscapes (in the presence of mutation, selection, and genetic drift) led to sustained diversity with semi-stable equilibria that depended on the ruggedness of the genotype-trait map. {Recent theoretical work \citep{mahadevan2025continual} on non-reciprocal ecological interactions (using only phenotype-fitness maps) has demonstrated the possibility of continual, steady-state evolution even without drift and in the weak mutation regime, but the role of genetic interactions in such scenarios remains unexplored. As of now,} no general theory exists on how intraspecific interactions alter the shape and the navigation of fitness landscapes.

Finally, explicit population dynamics, i.e., the consideration of population size change, may alter navigation on fitness landscapes. Here, extinction becomes a possible outcome of evolution on the landscape, such that fitness valleys could pose not only barriers to higher fitness but also dangers to the survival of an adapting population \citep{bell2017evolutionary,uecker2026modeling}. Moreover, the population size determines the mutational input to an evolving population, which, in turn, can alter the speed at which adaptation can proceed and the amount of (clonal) interference (see Section 4A). Whereas there exist several examples of studies that have considered constant population size as a parameter in models of fitness landscape navigation \citep{szendro2013predictability,li2024rapid}, we are not aware of fitness landscape theory that has systematically investigated the effect of population dynamics on the navigation of fitness landscapes.
\section{Discussion}

In this paper, we reviewed concepts, measures, and results pertaining to fitness landscape navigation, with a focus on models that map a discrete (hypercubic) genotype space to phenotypes and fitness. We highlight how both the topology and the topography of the fitness landscape affect how it is traversed by evolution. The term \textit{navigability} is commonly used in a regime in which a population evolves through beneficial or neutral single-mutational steps, and a substantial part of our paper (section 3) surveys the growing literature on landscape navigation in this field. Our review of the literature shows that no single measure of the shape of fitness landscapes serves as a satisfactory measure of navigability. To this end, we proposed a measure called {\em navigational ruggedness} which depends explicitly on average evolutionary outcomes, and only implicitly on detailed features of the landscape (through their effect on the evolutionary process). This allows for an improved comparison of the navigability of different fitness landscapes.


Whereas navigability is most commonly studied in the regime of adaptive walks on static landscapes, Section 4 of our review highlights examples of how the navigation process can change fundamentally in more realistic regimes beyond this standard assumption. Measures of navigability, on the other hand, have been developed mostly with a focus on accessible paths. However, navigability depends not only on the shape of the fitness landscape but also on the evolutionary forces under which the landscape is navigated. It remains for future work to examine the utility of existing navigability measures or to develop new measures applicable across diverse evolutionary regimes.

\section{Methods}

\subsection{Generating the landscapes}

The genotype space was generated by enumerating all $a^L$ genotypes as tuples of length $L$. Each locus could take $a$ integer values in the set $\mathcal{A} = \{0,1, \dots,a-1\}$, and the full genotype space was obtained as the Cartesian product of this set across all $L$ loci. For the landscapes in Figure \ref{fig:nr}, we used $L=4, a=4$.

\subsubsection*{HoC landscapes} The HoC landscapes were generated by assigning independent and identically distributed random variables drawn from the Normal distribution $\mathcal{N}(0,1)$ to each genotype in the genotype space. The landscape was linearly normalized to the interval $[0,1]$ by subtracting the minimum fitness and dividing by the fitness range.

\subsubsection*{NK landscapes} The NK landscapes were constructed following the standard NK model, where the fitness contribution of each locus depended on its own allele and those at $K$ interacting partner loci. For each locus $i \in \{1, \dots, L\}$, interaction partners were assigned deterministically as the next $K$ loci under periodic boundary conditions, ensuring a uniform interaction structure across the genome. Each locus was associated with a lookup table mapping allele combinations to fitness contributions, with values drawn independently from a standard normal distribution for each unique combination. Genotype fitness was computed as the sum of locus-specific contributions and subsequently normalized to $[0,1]$ by min–max scaling over the full landscape. We used $L=4, a=4$ and $K=1$ for the NK1 landscape and $K=2$ for the NK2 landscape in Figure \ref{fig:nr}.

\subsubsection*{RMF landscapes}

To generate the RMF landscapes, a reference genotype was chosen as the sequence consisting of identical alleles at all loci (e.g., the all-zero genotype). For each genotype $g$, the Hamming distance $d(g, g_{\mathrm{ref}})$ from the reference genotype $g_{\mathrm{ref}}$ was computed. The fitness of genotype $g$ was then defined as
\[
f(g) = -c \cdot \, d(g, g_{\mathrm{ref}}) + \eta(g),
\]
where $c$ is a parameter controlling the slope of the additive fitness gradient, and $\eta(g)$ is an independent random variable drawn from a standard normal distribution. After computing raw fitness values for all genotypes, the landscape was linearly normalized to the interval $[0,1]$ by subtracting the minimum fitness and dividing by the fitness range. We used $L=4, a=4$ and $c=1$ for the RMF1 landscape and $c=2$ for the RMF2 landscape in Figure \ref{fig:nr}.

\subsubsection*{Alternating landscapes} 

To generate the alternating landscapes, for each genotype $g$, the Hamming distance $d(g, g_{\mathrm{ref}})$ to the reference genotype was computed. Fitness was then assigned as an alternating function of this distance, such that genotypes at even Hamming distances were assigned higher fitness values, while those at odd distances were assigned lower fitness values. This results in nearly maximally rugged landscapes, similar to the egg-box model \citep{ferretti2016measuring} for $a=2$ or the model used by \cite{Rosenberg2005AdaptiveWalks} for $a>2$, but with some random fitness contribution added to each genotype.

Specifically, for even distances, fitness was defined as a weakly decreasing function of distance with added Gaussian noise,
\[
f(g) = 1 - 0.1\, \cdot d(g, g_{\mathrm{ref}}) + 0.01 \cdot \eta(g),
\]
where $\eta(g) \sim \mathcal{N}(0, 1)$. For odd distances, fitness values were assigned a lower baseline that decayed with distance,
\[
f(g) = \frac{0.1}{d(g, g_{\mathrm{ref}}) + 1} + 0.01 \cdot \eta(g),
\]
with $\eta(g)\sim \mathcal{N}(0, 1)$.

Fitness values were truncated to ensure non-negativity and subsequently normalized by dividing by the maximum fitness value in the landscape, such that the global peak fitness was rescaled to 1.

\subsection{Computing the ruggedness measures}

The ruggedness measures were computed as averages over 1000 independently generated landscapes of each type.

\subsubsection*{Navigational ruggedness} 
The navigational ruggedness $\mathcal{R}_N$ was computed by simulating $N$ random adaptive walks \citep{kauffman1987towards} on the fitness landscape, by starting from uniformly sampled random genotypes. Using this, each peak was associated with an empirical probability of reaching the peak $P_{\mathrm{reach}}(\text{peak})$, proportional to the number of random walks that terminated at that peak. $\mathcal{R}_N$ was then calculated as:
\[
\mathcal{R}_N = 1-\sum_{\text{peak}} P_{\mathrm{reach}}(\text{peak})\, f_{\text{rel}}(\text{peak}),
\]
where $f_{\text{rel}}(\text{peak})$ is the fitness of the peak relative to the global peak fitness. Here, we used $N=5000$.

\subsubsection*{Baseline ruggedness} 

The baseline ruggedness $\mathcal{R}_B$ was computed by calculating the average fitness of all $N_{\text{peaks}}$ peaks in the landscapes, i.e., $\bar{f} = \frac{1}{N_{\text{peaks}}} \cdot \sum_{\text{peak}} f_{\text{rel}}(\text{peak})$ and subtracting that from 1, to estimate how far the average fitness is from the global peak fitness.

\subsubsection*{Number of peaks} 

Fitness peaks were identified implicitly through the random adaptive walks on the landscape. Peaks are defined as genotypes with no fitness-increasing mutations, and were thus detected as absorbing states of the random adaptive walk dynamics.

\subsubsection*{Number of non-global peak basin genotypes} 

A directed graph was constructed in which an edge from genotype $g$ to genotype $n$ exists if $n$ is a single-mutant neighbor of $g$ with higher fitness. This graph was subsequently reversed so that edges point from higher-fitness genotypes to lower-fitness ones. Then, starting from the global peak, a depth-first search was performed on the reversed graph to identify all genotypes that can reach the peak through monotonic fitness-increasing paths in the original landscape. The number of visited genotypes defines the size of the global peak basin. All remaining genotypes were classified as non-global-peak genotypes.

\setcounter{figure}{0}
\renewcommand{\figurename}{Supplementary Figure}

\begin{figure*}[htbp]
\centering
\includegraphics[width=500pt]{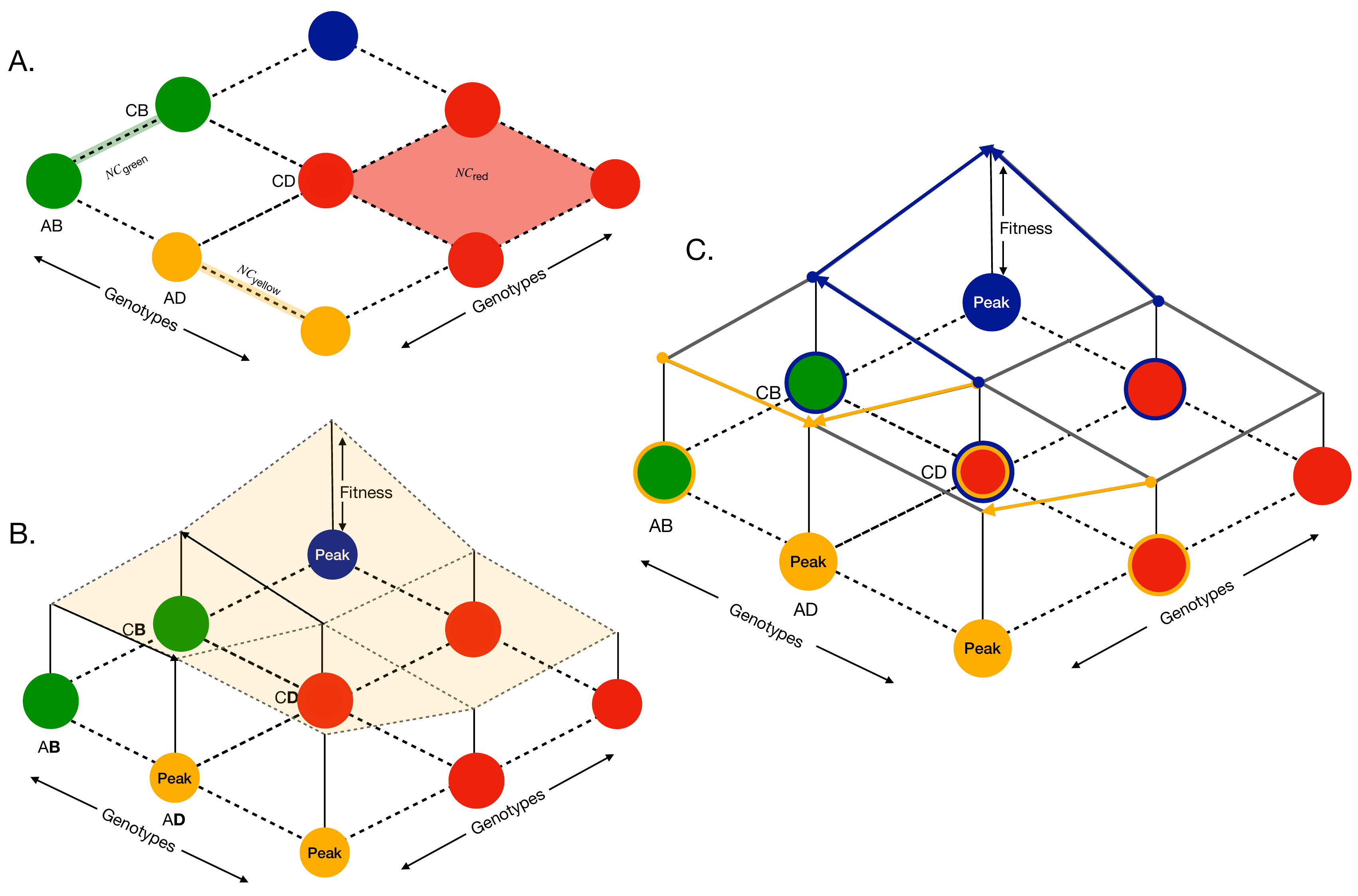}
\caption{\textbf{Schematic diagrams of a GP and a GPF map along with dynamics on the map.} \textbf{A.} Circles represent genotypes, and edges connect genotypes to their \textit{genotypic neighbors}. The colour of each genotype represents its phenotype; together, the circles and their colours define a \textit{genotype--phenotype (GP) map}. The red genotype $CD$ has a \textit{genotypic robustness} of $\frac{1}{2}$. The genotypes at the vertices of the red square form the largest \textit{neutral component (NC)}, of size 4. The \textit{phenotypic evolvability} of the NC corresponding to the red phenotype, $\epsilon_{\text{red}}$, is 3, since three distinct phenotypes (blue, green, and yellow) can be reached by single mutational steps from the red NC. Similarly, $\epsilon_{\text{green}} = 3$, $\epsilon_{\text{yellow}} = 2$, and $\epsilon_{\text{blue}} = 2$.  \textbf{B.} The vertical elevation of each genotype corresponds to its fitness, and these elevations collectively define the genotype--fitness map, or the \textit{fitness landscape}. The genotypes $AB$, $AD$, $CD$, and $CB$ exhibit \textit{sign epistasis}, since the mutation from $B \to D$ is beneficial on the $A$ background but deleterious on the $C$ background (shown with thin black arrows). \textbf{C.} Dynamics on the landscape: 
 Coloured arrows indicate \textit{accessible paths}: they point in the direction of increasing fitness, and their colour denotes the peak phenotype at which the walk terminates. The \textit{adaptive basin} of each peak (here blue and yellow) is highlighted by the border colour of the genotypes belonging to that basin. Neutral mutations are shown as dark grey undirected edges. In a greedy walk, a population at this genotype proceeds to the yellow peak genotype $AD$. However, in a random \textit{adaptive walk}, the population may instead move to the green genotype $CB$ and ultimately reach the blue peak.   
}\label{sfig:1}
\end{figure*}

\begin{figure*}[htbp]
\centering
\includegraphics[width=500pt]{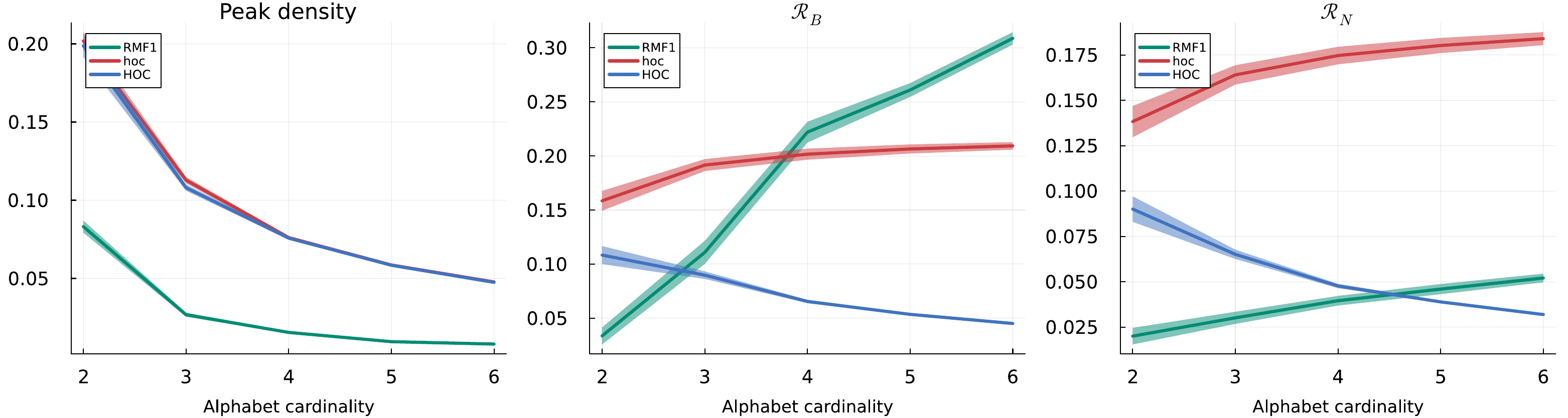}
\caption{\textbf{Effect of landscape dimensionality on ruggedness.} The density of peaks (fraction of genotypes that are local fitness peaks) decreases with increasing alphabet cardinality for all three landscape classes. In contrast, the response of navigational ruggedness to increasing dimensionality is governed by changes in baseline ruggedness, reflecting how the mean fitness of peaks varies with alphabet cardinality. Because baseline ruggedness depends on the underlying distribution of fitness values, both baseline and navigational ruggedness exhibit distribution-dependent trends. Consequently, HoC landscapes generated from a normal distribution (red lines) and a uniform distribution (blue lines) show qualitatively different responses to increasing alphabet cardinality. RMF landscapes (green lines) exhibit an increase in navigational ruggedness with increasing alphabet cardinality.}\label{sfig:2}
\end{figure*}

\section{Code availability}

Code used to produce Figure~\ref{fig:nr} is available 
\href{https://github.com/MSri95/Landscape_Navigation/blob/main/Landscape_Navigation.ipynb}{here} and will be archived on Zenodo after acceptance of the paper.

\section{Acknowledgments}
JK is grateful to Daniel Oros for useful discussions, and to the MPI for Evolutionary Biology (Pl\"on) and the National Institute for Theory and Mathematics in Biology (Chicago) for hospitality during the completion of the manuscript.
During the preparation of this manuscript, the authors used the generative AI tools ChatGPT and Claude for aid in literature search and for editorial purposes.

\section{Funding}
JK acknowledges support by Deutsche Forschungsgemeinschaft (DFG) within projects CRC 1310 \textit{Predictability in evolution} and TRR 341 \textit{Plant ecological genetics}. SGD and CB are grateful for support from SNF Project Grant 320030-236379 from the Swiss National Science Foundation to CB.

\section{Conflicts of interest}
The authors declare that they have no conflicts of interest.

\bibliographystyle{plainnat}
\bibliography{Nav_bibliography_clean}

\end{document}